\documentclass[twocolumn,amsmath,amssymb,nofootinbib,a4paper,preprintnumbers]{revtex4} 

\usepackage{tikz}
\usepackage[utf8]{inputenc}
\usepackage{graphicx} 
\usepackage{amsmath}
\usepackage{amsfonts,amsbsy}
\usepackage{amssymb}
\usepackage{subdepth}
\usepackage{bbold}

\usepackage{xcolor}
\definecolor{lcolor}{rgb}{0.5,0,0}
\definecolor{citcolor}{rgb}{0,0.3,0.0}
\usepackage[breaklinks,colorlinks,urlcolor=blue,citecolor=citcolor,linkcolor=lcolor]{hyperref}

\newcommand{\xx}{{\mathbf{x}}}

\newcommand{\ii}{{\boldsymbol{\hat{\i}}}}
\newcommand{\jj}{{\boldsymbol{\hat{\j}}}}

\newcommand{\ptt}{{p_T}}

\newcommand{\ud}{\mathrm{d}}

\newcommand{\tr}{\, \mathrm{Tr} \, }
\newcommand{\R}{\mathrm{Re}}
\newcommand{\nc}{{N_\mathrm{c}}}

\newcommand{\half}{\frac{1}{2}}
\newcommand{\ah}{\mathrm{ah}}

\newcommand{\qs}{Q_{\mathrm{s}}}

\newcommand{\lqcd}{\Lambda_{\mathrm{QCD}}}

\newcommand{\nr}[1]{(\ref{#1})} 

\newcommand{\fig}{Fig.~}

\newcommand{\eq}{Eq.~}

\newcommand{\eqs}{Eqs.~}

\begin{document}

\title{
Time evolution of linearized gauge field fluctuations on a real-time lattice
}
\preprint{CERN-TH-2016-208}

\author{A. Kurkela}
\email{a.k@cern.ch}
\affiliation{
Theoretical Physics Department, CERN, Geneva, Switzerland
}
\affiliation{Faculty of Science and Technology, University of Stavanger, 4036 Stavanger, Norway}

\author{T. Lappi}
\email{tuomas.v.v.lappi@jyu.fi}
\affiliation{
Department of Physics, %
 P.O. Box 35, 40014 University of Jyv\"askyl\"a, Finland
}
\affiliation{
Helsinki Institute of Physics, P.O. Box 64, 00014 University of Helsinki,
Finland
}

\author{J. Peuron}
\email{jarkko.t.peuron@student.jyu.fi}
\affiliation{
Department of Physics, %
 P.O. Box 35, 40014 University of Jyv\"askyl\"a, Finland
}

\begin{abstract}
Classical real-time lattice simulations play an important role in understanding non-equilibrium phenomena in gauge theories and are used in particular to model the prethermal evolution of heavy-ion collisions. 
Due to instabilities, small quantum fluctuations on top of the classical background 
may significantly affect the dynamics of the system.
In this paper we argue for the need for a numerical calculation of a system of classical gauge fields and small linearized fluctuations in a way that keeps the separation between the two manifest. We derive and test an explicit algorithm to solve these equations on the lattice, maintaining gauge invariance and Gauss's law.
\end{abstract}

\maketitle

\section{Introduction}

Particle production at central rapidities in collisions of high energy hadrons or nuclei is dominated by the clouds of small-$x$ gluons surrounding the projectiles. The high density of these gluons has been argued to lead to ``gluon saturation'', i.e., the emergence of a dominant semihard transverse momentum scale $\qs \gg \lqcd$ where the physics become nonperturbative due to the nonlinear interactions of the gluons even at weak coupling ~\cite{Gelis:2010nm}.
The saturation picture of a weak coupling and a nonperturbatively large phase space density of gluons $f\sim 1/g^2$ leads to a description of the initial stages of a heavy ion collision in terms of ``glasma'' fields~\cite{Lappi:2006fp}, strong boost invariant color fields with transverse coherence length $\sim \qs^{-1}$ . How these maximally anisotropic far-from-equilibrium gauge fields
hydrodynamize, isotropize, and reach local thermal equilibrium to form quark-gluon plasma has been a central open question in understanding the spacetime evolution of the matter produced in a heavy-ion collisions.

The large phase space occupancy, or equivalently the strength of the gauge fields, at the early stages of the collision admits a classical description of the 
glasma fields accurate to leading order in $g$. 
The classical description, however, poses a problem phenomenologically as the boost invariance of the fields is not broken and the system remains anisotropic at all times, never thermalizing or reaching hydrodynamical flow.

For the process of isotropization to proceed, it is necessary (but not sufficient) that the boost invariance is broken by small rapidity-dependent  fluctuations. The origin of the fluctuations may be quantum \cite{Fukushima:2006ax,Dusling:2010rm,Epelbaum:2011pc,Dusling:2012ig,Epelbaum:2013waa} 
or arise from the longitudinal structure of the colliding nuclei \cite{Gelfand:2016yho, Schenke:2016ksl}. It is then expected that in the presence of the anisotropic background, some of these  fluctuations are unstable and experience a period of exponential growth, playing an important role in the isotropization process~\cite{Mrowczynski:1994xv,Mrowczynski:1996vh,Mrowczynski:2004kv,Kurkela:2011ub}.

Assuming a parametric scale separation between the dominant scale $\qs$ and the inverse wavelength of the unstable modes $g f^{1/2} \qs$, the growth and saturation of the plasma instabilities can be studied in a ``hard loop'' (HL) framework in which the modes at the scale $\qs$ are treated as quasiparticles and the unstable modes as classical fields. Many calculations have been performed in this framework both analytically~\cite{Arnold:2003rq,Romatschke:2003ms,Romatschke:2004jh,Arnold:2004ti,Arnold:2004ih,Kurkela:2011ub,Rebhan:2005re,Rebhan:2004ur,Kurkela:2011ti}  and numerically~\cite{Nara:2005fr,Dumitru:2005gp,Bodeker:2007fw,Rebhan:2008uj,Attems:2012js}. This is indeed a valid approach when the isotropization process is already under way and the system is only moderately anisotropic and the occupation numbers $f$ of gluonic states with $\ptt \sim \qs$ have decreased from their initial value $\sim g^{-2}$. The method however fails at the earliest time scale after the collision, $\tau \sim 1/\qs$, when the role of the instabilities are expected to be the most important.

The contribution of plasma instabilities to isotropization has also been studied using purely classical field simulations~\cite{Romatschke:2005pm,Romatschke:2006nk,Berges:2012cj,Berges:2013eia} without performing the Hard Loop approximation. These calculations typically proceed using the so called ``classical statistical approximation'' (CSA). This consists of identifying the initial field fluctuations of the fields, adding these to the classical background field, and then solving the time evolution of the system using the full classical equations of motion on a discrete lattice. Some of these calculations have pointed towards the possibility of a very rapid isotropization caused by the plasma instabilities seeded by the quantum fluctuations of the gauge fields~\cite{Gelis:2013rba}.  

The treatment of quantum fluctuations in CSA however is problematic due the backreaction of the fluctuations on the background field. Including the quantum fluctuations in the equations of motion of the background is justified only for the modes that
grow large and become effectively classical \cite{Khlebnikov:1996wr,Khlebnikov:1996mc}; for the other modes, the time evolution of the fluctuations is mistreated. The problem is severe in the case of quantum fluctuations, which have a highly UV-divergent spectrum and occupy modes with $f\sim 1/2$ at all scales supported by the lattice. In the CSA these fluctuations are superimposed on top of the background with $f\sim 1/g^2$  \cite{Gelis:2013rba}. Even though the occupancy of the mistreated fluctuations is parametrically smaller than that of the background field, the phase space opens up like $\int^{1/a} \ud^3 p$, with lattice spacing $a$. Therefore, on a fine enough lattice the UV tail of the fluctuation spectrum dominates the energy density, particle number, and eventually the dynamics of the system%
\footnote{ Note that a gauge theory (unlike scalar theory studied in the cosmological context) is particularly sensitive to UV modes as the inelastic collisions of the modes 
can rapidly move the energy towards the IR.
} 
and the time evolution of the combined system cannot be reliably followed in a classical simulation \cite{Moore:2001zf}. No continuum limit may be taken (see also \cite{Epelbaum:2014yja}).

To avoid this problem, we propose to study the evolution of the fluctuations on a mode-by-mode basis in a 
setup where the evolution of the fluctuation is explicitly linearized.  In this case one can treat the fluctuations to one-loop order, explicitly excluding interactions between the fluctuations and any backreaction to the classical field. One loses the ability to resum late-time ``secular divergences'' that was one of the motivations for adopting the CSA~\cite{Dusling:2010rm,Dusling:2012ig}. However, the later-time behavior and eventual hydrodynamization in the context of a heavy ion collision is in any case better described in terms of kinetic theory \cite{Epelbaum:2015vxa,Kurkela:2015qoa,Keegan:2016cpi}. 
Instead, one keeps the analytical control given by a well defined weak coupling expansion, where different orders in $g$ remain separate. 
The growth and evolution of the unstable modes can be followed in a clean numerical setup, and one may choose to include only the unstable modes in the simulation. 
 One can also formulate the  calculation of gluon production in a dense-dense collision system to NLO accuracy~\cite{Gelis:2008rw,Gelis:2008ad} analogously to the way quark pair production from the classical field is calculated by solving the Dirac equation in the classical background~\cite{Gelis:2003vh,Gelis:2004jp,Gelis:2005pb,Gelfand:2016prm,Mueller:2016ven}.

We will write down the equations of motion for the system of a classical gauge field and linearized fluctuations in Sec.~\ref{sec:eoms}, noting in particular that maintaining Gauss's law in a calculation with discretized time requires some care. In Sec.~\ref{sec:num} we will present results from simple numerical tests of our algorithm, before pointing in Sec.~\ref{sec:conc} towards some of its potential future applications.

\section{Equations of motion for fluctuations}
\label{sec:eoms}
In this Section we construct the equations of motion for the linearized fluctuations of the gauge and the chromoelectric field $\{a_i, e^i\}$ on top of the background field. 
On the lattice we will use the Kogut-Susskind Hamiltonian~\cite{Kogut:1974ag} for the background field and in discretizing the equations 
of motion for the background field we will take special care to make sure that the discretized and linenarized equations
of motion exactly conserve the Gauss's law constraint. 

In this paper, for simplicity, we will constrain the discussion to a system not undergoing longitudinal expansion (fixed box), however, the extension to a expanding coordinate system is trivial.

\subsection{Small fluctuations in the continuum}

In the continuum the Hamiltonian of a pure gauge theory can be written, in temporal gauge $A_0=0$ as
\begin{equation}
 H =\int \ud^3\xx \left[ \tr E^i E^i + \half \tr F_{ij} F_{ij}\right], 
\end{equation}
with field strength tensor $F_{ij} = (ig)^{-1}[D_i,D_j] = \partial_i A_j - \partial_j A_i + i g [A_i,A_j]$, where the covariant derivative is $D_i = \partial_i +igA_i$. Here we write the gauge and chromoelectric fields in matrix form $ A_i = A_i^a t^a,$ with the fundamental representation generators $t^a$ normalized as $\tr t^at^b = \half \delta^{ab}$. From this Hamiltonian one derives the equations of motion
\begin{eqnarray}
\dot{A}_i  &= &E^i
\\
\dot{E}^i  &= & \left[ D_j,F_{ji} \right].
\end{eqnarray}
In order to project to the physical charge sector, also the Gauss's law constraint must be fulfilled
\begin{equation}
 C(\xx,t) \equiv \left[  D_i , E^i \right] =0,
\end{equation}
which is conserved exactly by the equations of motion $\partial_t C(\xx,t) = 0$.

Dividing the field into a background field and linearized fluctuations
\begin{equation}
 (E^i,A_i) \to (E^i  + e^i,  A_i + a_i),
 \end{equation} the equations of motion and Gauss's law for the fluctuations become
\begin{eqnarray}
\label{eq:contaeom}
 \dot{a}_i &= &e^i 
\\
\dot{e}^i &=& \left[D_j,\left[D_j,a_i\right]\right] 
- \left[D_j,\left[D_i,a_j\right]\right] + i g \left[a_j,F_{ji}\right]
\\ &=&
\label{eq:conteeom}
\left[D_j,\left[D_j,a_i\right]\right] 
- \left[D_i,\left[D_j,a_j\right]\right] + 2 i  g \left[a_j,F_{ji}\right]
\end{eqnarray}
where the second form of the equation for $\dot{e}^i$ allows for an interpretation in the background field gauge $[D_i,a^i]=0$ in terms of an adjoint representation scalar field equation for $a_i$ supplemented with a gluon chromomagnetic moment term (see, e.g., \cite{Greiner:1985ce}) . Similarly, the Gauss's law constraint for the fluctuation reads
\begin{equation}
\label{eq:contgauss}
c(\xx,t) =\left[ D_i,e^i \right]  + i g \left[a_i,E^i\right] = 0.
\end{equation}

\subsection{Discretized equations for background}

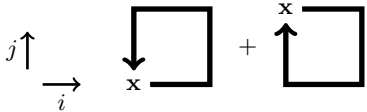
\begin{figure}
 \begin{tikzpicture}
 \draw[->, line width=1pt] (-1.4,0.2) -- node[left] {$j$}(-1.4,0.7);
\draw[->, line width=1pt] (-1.2,0) -- node[below] {$i$}  (-0.7,-0.);
\node (base) at (0,0) {$\xx$};
  \draw[->, line width=2pt] (base)  -- ++(1,0) -- ++(0,1) -- ++(-1,0) -- (base);
\node (base2) at (2,1) {$\xx$};
  \draw[->, line width=2pt] (base2)  -- ++(1,0) -- ++(0,-1) -- ++(-1,0) -- (base2);
 \node at(1.5,0.5) {+};
\end{tikzpicture}
\caption{Plaquettes for the timestep of the electric field $\Box_{ij}(\xx) + \Box_{i,-j}(\xx)$.}\label{fig:clestep}
\end{figure}

In order to conserve the gauge symmetry exactly, it is convenient to 
trade the gauge fields $A_i$ belonging to the Lie algebra of the group 
to link matrices $U_i$ which are members of the group. 
The Kogut-Susskind Hamiltonian~\cite{Kogut:1974ag} in terms of the link matrices reads
\begin{multline}\label{eq:ksh}
H =\frac{a^3}{g^2} \sum_\xx \Bigg\{ \tr \big[ a^{-2} E^i(\xx) E^i(\xx) \big]
\\
+ \frac{2}{ a^4} \sum_{i<j} \R\tr\big[ \mathbb{1}  -  \Box_{i,j}(\xx) \big]
  \Bigg\},
\end{multline}
where the spatial coordinate ${\bf x}$ takes discrete values on a cartesian lattice ${\bf x}=a (n_i,n_j,n_k)$, with integers $n_i,n_j,n_k$ and lattice spacing $a$.
Here the plaquette $\Box_{i,j}(\xx)$ is written in terms of the link matrices 
$U_i(\xx)$ 
\begin{equation}
 \Box_{i,j}(\xx) = U_i(\xx) U_j(\xx+\ii) U^\dag_i(\xx+\jj)U^\dag_j(\xx),
\end{equation}
where $\ii,\jj$ are unit vectors in the $i,j$ directions;
see \fig\ref{fig:clestep} for an illustration.%
\footnote{Note that in the discrete formulation from now on we abandon the summation convention for spatial indices $i,j,\dots$ (but not for color indices).}
The lattice fields are related to continuum quantities by
$U_i(\xx)\approx e^{i a g A_i(\xx)}$
and  $E^i_\textup{lat} \approx a g E^i_\textup{cont}$.

The Kogut-Susskind Hamiltonian gives us equations of motion that are discrete in space but continuous in time:
\begin{eqnarray}
\dot{U}_i(\xx) &=& i E^i(\xx) U_i(\xx) 
\\
a^2 \dot{E}^i(\xx) &=& - \sum_{j\neq i } \left[\Box_{i,j}(\xx) + \Box_{i,-j}(\xx) \right]_\ah, \label{eq:Edot}
\end{eqnarray}
where the plaquette in the negative $j$ direction is
 $\Box_{i,-j}(\xx) =  U_i(\xx) U^\dag_j(\xx+\ii-\jj) U^\dag_i(\xx-\jj)U_j(\xx-\jj)$. Here the notation  $[]_\ah$ denotes the antihermitian traceless part of a  matrix:
\begin{equation}
[V]_\ah \equiv \frac{-i}{2} \left[ V - V^\dag - \frac{\mathbb{1}}{\nc}\tr(V-V^\dag)\right],
\end{equation}
where $\nc$ is the number of colors.

In order to perform a practical simulations, also the time direction must be discretized.
To guarantee time reversal invariance and second order accuracy in the time step $\ud t$, the time direction is commonly discretized with the leaprog algorithm, where the electric fields and the links live on alternate timesteps
\begin{align}
 U(t+\ud t) &= e^{i E^i(t+\ud t/2) \ud t} U_i(t)
\label{eq:cllinkstep}
\\
\label{eq:clestep}
a^2 E^i(t+\ud t) &=a^2  E^i(t)  \\  -\ud t \sum_{j\neq i } & \left[\Box_{i,j}\left(t+\frac{\ud t}{2}\right) + \Box_{i,-j}\left(t+\frac{\ud t}{2} \right) \right]_\ah, \nonumber
\end{align}
 where we have dropped the explicit position arguments for brevity.
 It is a straightforward exercise to show that both the link and electric field timesteps \nr{eq:cllinkstep} and \nr{eq:clestep} separately conserve the discretized version of Gauss's law constraint
\begin{align}\label{eq:clgauss}
&C(\xx,t) = \sum_i \frac{1}{a^2}\left\{E^i(\xx) - U^\dag_i(\xx-\ii)E^i(\xx-\ii)U_i(\xx-\ii)\right\}, \nonumber\\
&C(\xx,t+\ud t)= C(t).
\end{align}
Finally, let us recall that under a lattice gauge transformation $V(\xx)$ (which must be time-independent in order to conserve the temporal gauge condition) the links and electric fields transform as
\begin{eqnarray}
\label{eq:gaugetrlink}
U_i(\xx) &\to& V(\xx) U_i(\xx)V^\dag(\xx+\ii) 
\\
\label{eq:gaugetre}
 E^i(\xx) &\to& V(\xx) E^i(\xx)V^\dag(\xx).
\end{eqnarray}
It is easy to see that the Hamiltonian~\nr{eq:ksh} is gauge invariant and the equations of motion~\nr{eq:cllinkstep},~\nr{eq:clestep} and Gauss's law~\nr{eq:clgauss} gauge covariant under these transformations.

\begin{figure}
 \begin{tikzpicture}
\draw[->, line width=1pt] (-1.4,0) -- node[left] {$j$}(-1.4,0.5);
\draw[->, line width=1pt] (-1.2,-0.2) -- node[below] {$i$}  (-0.7,-0.2);
\node(b1)  [line width=1pt,circle,draw,inner sep=0.2pt] at (0,0){$\boldsymbol{\rightarrow}$};
\draw [line width=1.5pt,->] (b1) -- ++(1,0) -- ++(0,1) -- ++ (-1,0) -- ++(0,-0.6); 
\node at (1.5,0.5) {$+$};
\node(b2) at (2,0){};
\node(b2b)  [line width=1pt,circle,draw,inner sep=1pt] at (3,0){$\boldsymbol{\uparrow}$};
\draw [line width=1.5pt,->] (b2) -- (b2b) --  ++(0,1) -- ++ (-1,0) -- ++(0,-0.8); 
\node at (3.5,0.5) {$-$};
\node(b3) at (4,0){};
\node(b3b)  [line width=1pt,circle,draw,inner sep=0.2pt] at (4,1){$\boldsymbol{\rightarrow}$};
\draw [line width=1.5pt,->] (b3) -- ++(1,0) -- ++(0,1) -- (b3b)  -- ++(0,-0.8); 
\node at (5.5,0.5) {$-$};
\node(b4)  [line width=1pt,circle,draw,inner sep=1pt] at (6,0){$\boldsymbol{\uparrow}$};
\draw [line width=1.5pt,->] (6.4,0) -- ++(0.6,0) -- ++(0,1) -- ++ (-1,0) -- ++(b4); 
\node at (-0.5,-1.5) {$+$};
\node(b5)  [line width=1pt,circle,draw,inner sep=0.2pt] at (0,-1){$\boldsymbol{\rightarrow}$};
\draw [line width=1.5pt,->] (b5) -- ++(1,0) -- ++(0,-1) -- ++ (-1,0) -- ++(0,0.6); 
\node at (1.5,-1.5) {$-$};
\node(b6) at (2,-1){};
\node(b6b)  [line width=1pt,circle,draw,inner sep=1pt] at (3,-2){$\boldsymbol{\uparrow}$};
\draw [line width=1.5pt,->] (b6) --  ++(1,0) -- (b6b) -- ++ (-1,0) -- ++(0,0.8); 
\node at (3.5,-1.5) {$-$};
\node(b7) at (4,-1){};
\node(b7b)  [line width=1pt,circle,draw,inner sep=0.2pt] at (4,-2){$\boldsymbol{\rightarrow}$};
\draw [line width=1.5pt,->] (b7) --  ++(1,0) -- ++(0,-1) -- (b7b)  -- ++(0,0.8); 
\node at (5.5,-1.5) {$+$};
\node(b8) at (6,-1){};
\node(b8b)  [line width=1pt,circle,draw,inner sep=1pt] at (6,-2){$\boldsymbol{\uparrow}$};
\draw [line width=1.5pt,->] (b8) --  ++(1,0) -- ++(0,-1) -- (b8b)  -- ++(0,0.8); 
\end{tikzpicture}
\caption{Plaquettes for the timestep of electric field fluctuation, \eq\nr{eq:estep}. The circled arrows in directions $i,j$ denote the field fluctuation $a_i$, $a_j$. The solid lines are link matrices, with a gap at position $\xx$ where the expression gauge transforms. Note that if the circle is next to the gap ($a_i$ at position $\xx$), the gap can be on either side of the circle, corresponding to $a_i\Box$ or $\Box a_i$. 
The ordering of the terms is the same as in \eq\nr{eq:estep}.
}\label{fig:estep}
\end{figure}
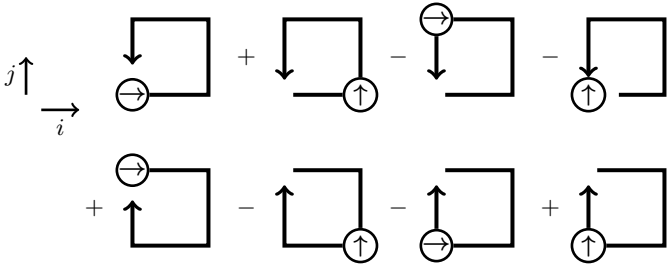

\subsection{Discretized equations for fluctuations}

After these preliminaries, let us move to the lattice equations of motion for the small fluctuations. 
Naturally, there is a certain freedom in writing down the discretized equations; here, we choose to construct the discretized equations so that they satisfy the following requirements:
\begin{enumerate}
 \item Reduction to the continuum equations of motion \nr{eq:contaeom},~\nr{eq:conteeom} in the limit $a \to 0, \  \ud t \to 0$.
\item \label{it:gt} Gauge covariance under the transformations \nr{eq:gaugetrlink},~\nr{eq:gaugetre}.
 \item Linearity in $a_i$ and $e^i$.
\item An exact conservation of a lattice version of a Gauss's law that reduces to \nr{eq:contgauss} in the limit $a \to 0, \ \ud t\to 0$ at every time step.
\item Time reversal invariance (under $\ud t \to -\ud t$).
\end{enumerate}
We choose here to start from condition \ref{it:gt} by defining the required gauge transformation properties as those of an adjoint representation scalar field:
\begin{eqnarray}
 a_i(\xx) &\to& V(\xx) a_i(\xx) V^\dag(\xx)
\\
 e^i(\xx) &\to& V(\xx) e^i(\xx) V^\dag(\xx).
\end{eqnarray}
From these it follows that $a_i$ must correspond to a variation of the link  matrix $U_i(\xx)$ on the left:
\begin{equation}
 U_i(\xx)_\textup{bkg + fluct} = e^{i a_i(\xx)}U_i(\xx) \approx
U_i(\xx) + i a_i(\xx)U_i(\xx).
\end{equation}
In the continuum limit, the fluctuation field on the lattice is related to the continuum equivalent though $a_i^\textnormal{lat} = a g a_i^\textnormal{cont}$. 

We then \emph{choose} to discretize the perturbation of the electric field by linearizing the r.h.s. of 
\nr{eq:clestep}, so that 
\begin{widetext}
\begin{align} \label{eq:estep}
a^2 e^i(t+\ud t) = &a^2 e^i(t) - \ud t \sum_{j\neq i}\Bigg[ i \Big(
 a_i(\xx)\Box_{i,j}(\xx) 
+ a_j(\xx+\ii \to \xx)\Box_{i,j}(\xx)
 - \Box_{i,j}(\xx)a_i(\xx+\jj \to \xx) 
- \Box_{i,j}(\xx)a_j(\xx) 
\\
& + a_i(\xx)\Box_{i,-j}(\xx) 
 - a_j(\xx+\ii-\jj \to \xx+\ii \to \xx)\Box_{i,-j}(\xx) 
 - \Box_{i,-j}(\xx) a_i(\xx-\jj \to \xx)
+\Box_{i,-j}(\xx)a_j(\xx-\jj\to \xx)
 \Big) \Bigg]_\ah, \nonumber
\end{align}
\end{widetext}
which is easily seen to be gauge covariant. Figure~\ref{fig:estep} illustrates the ordering of the plaquettes and the field fluctuations in \eq\nr{eq:estep}.
Here we denote by the fluctuation parallel transported from site $\xx+\ii$ to site $\xx$ by
\begin{equation}
a_j(\xx+\ii \to \xx)
 \equiv
U_i(\xx)a_j(\xx+\ii)U^\dag_i(\xx),
\end{equation}
and similarly for the fields parallel transported over two links%
\footnote{Note that, in our notation there are two identical ways of writing the most complicated terms involving
parallel transports over two links 
\begin{align}
& a_j(\xx+\ii-\jj \to \xx+\ii \to \xx)\Box_{i,-j}(\xx)
\\ \nonumber & \quad 
= 
\Box_{i,-j}(\xx)a_j(\xx+\ii-\jj \to \xx-\jj \to \xx).
\end{align}
}
\begin{align}
a_j(\xx+\ii-\jj \to \xx+\ii & \to \xx)  \\
\equiv U_i(\xx)&a_j(\xx+\ii-\jj \to \xx+\ii)U^\dag_i(\xx)\nonumber,
\end{align}
and so on. The links and gauge field fluctuations $a_i$ in \nr{eq:estep} are evaluated according to the leapfrog scheme at time $t+\ud t/2$.
We emphasize that the choice \nr{eq:estep} is not unique, but one could add terms proportional to $(\ud t)^2$ or higher powers. 

Similarly to the timestep of $E^i$, Gauss's law \nr{eq:clgauss} is linear in the chromoelectric field. The natural choice is then to derive Gauss's law for the fluctuations by replacing $E^i$ with $E^i + e^i$, $U_i(\xx)$ with
$U_i(\xx) + i a_i(\xx)U_i(\xx)$, and taking the linear terms in the fluctuation fields. This yields 
\begin{align} \label{eq:gauss}
 c(\xx&,t)
= \sum_i \frac{1}{a^2} \Big\{ e^i(\xx)- U_i^\dagger(\xx-\ii) e^i(\xx-\ii) U_i(\xx-\ii)\nonumber
 \\
&+ i U_i^\dagger(\xx-\ii)[a_i(\xx-\ii),E^i(\xx-\ii)]U_i(\xx-\ii) \Big\} .
\end{align}
We now have equations for the timestep of the electric field fluctuation $e^i$ and Gauss's law for the fluctuations. To complete the set of equations, we need also to specify the timestep for $a_i(\xx)$. The first guess would be a straightforward discretization of the continuum $\dot{a}_i = e^i$. However, this naive discretization is inadmissible, since it does not conserve the linearized Gauss's law \nr{eq:gauss}. Physically this would mean an unphysical creation of ``charges'' in the lattice. This can be traced to the fact that Gauss's law involves a covariant derivative using links $U_i$ that advance in time simultaneously as their fluctuations $a_i$, and the timestep must reflect this change.  Another hint of the subtlety of the step for $a_i$  is to see that a linearization of the timestep for the link $U_i(\xx)$ in \eq\nr{eq:cllinkstep} would involve developing the exponential $e^{i(E^i+e^i)\ud t}$ to linear order in $e^i$, which is a rather complicated expression when $E^i$ and $e^i$ do not commute.

We may, however, construct a valid update for the gauge fields by demanding the Gauss's law constraint to be conserved,
\begin{align}
c(\xx,t)  = c(\xx,t+\ud t).
\end{align}
It is straightforward to see that this condition holds if the update satisfies\footnote{ We drop the explicit time argument for the electric field from now on; this will always be dictated by the leapfrog scheme.}
\begin{equation}\label{eq:genstep}
\left[ E^i,a_i(t+\ud t) \right] = -i \left(\Box_{0i}e^i \Box^\dagger_{0i} - e^i\right)
+
\left[E^i,\Box_{0i} a_i(t)\Box^\dagger_{0i}\right],
\end{equation}%
where we use a shorthand for the ``timelike plaquette'' $\Box_{0i}=e^{i E^i\ud t}$. Imposing this condition on the 
gauge field update leads by construction to a time step that conserves Gauss's law. 

It is convenient here to separate the parts of $a_i,e^i$ that are parallel and perpendicular to 
$E^i$ in color space. Denoting
\begin{eqnarray}
 f^\parallel  &=& \frac{\tr \left[f E^i\right]}{\tr \left[ E^i E^i\right]}E^i,
\\
 f^\perp  &=& f - \frac{\tr \left[f E^i\right]}{\tr \left[ E^i E^i\right]}E^i,
\end{eqnarray}
\eq\nr{eq:genstep} can be solved for $a^\perp_i(t+\ud t)$
 in terms of $a^\perp_i(t)$ and $e^{i \perp}$ giving the equation of motion
 for the perpendicular component. 
Because of the commutator, Eq. \nr{eq:genstep} gives no condition for the parallel 
component, and we may complete the equations of motion with the naive discretization 
\begin{equation}
a^\parallel_i(t+\ud t)=a^\parallel_i(t) + \ud t e^{i \parallel }(t+\ud t/2)
\end{equation}
that already satisfies \eq\nr{eq:genstep}.

For a practical algorithm, it still remains a technical problem to solve the 
perpendicular components of the gauge field fluctuations from Eq.~\nr{eq:genstep}.
For a general gauge group we can write the solution more compactly in the adjoint representation: in terms of  the unitary matrix $\left(\widetilde{\Box}_{0i}\right)^{ab} = 2 \tr \left[t_a \Box_{0i} t_b \Box_{0i}^\dag \right]$, the hermitian matrix $\left(\widetilde{E}^i\right)^{ab} = E^i_c \left(T^c\right)^{ab}
= -if_{cab}E^i_c$ and the $N_c^2-1$ component vectors $\underline{a}_i$ and $\underline{e}^i$ with components $\left(a_i\right)^a$ and $(e^i)^a$. In this notation \eq\nr{eq:genstep} becomes
\begin{equation}
 \widetilde{E}^i \underline{a}_i(t+\ud t) = -
i \left( \widetilde{\Box}_{0i} - \mathbb{1} \right) \underline{e}^i
+ \widetilde{E}  \widetilde{\Box}_{0i}\underline{a}_i(t)
\end{equation}
The parallel components are the null space of the matrix $\widetilde{E}^i$, and in this subspace the timelike plaquette acts like the identity: $\widetilde{\Box}^\dag_{0i} \underline{f}^\parallel = \underline{f}^\parallel$. Thus parallel components of $\underline{a}^i(t)$ and $\underline{e}_i$ only generate parallel components of $\underline{a}^i(t+\ud t)$. In the perpendicular color directions, on the other hand, the matrix $\widetilde{E}^i$  is invertible, and we can write the gauge field timestep as
\begin{multline}\label{eq:finalastep}
 \underline{a}_i(t+\ud t) = \left(\widetilde{E}^i\right)^{-1}_{\perp}\left[
-i \left( \widetilde{\Box}_{0i} - \mathbb{1} \right) \underline{e}^{i \perp}
+ \widetilde{E}^i  \widetilde{\Box}_{0i}\underline{a}_i^\perp(t)
\right]
\\
+ 
\underline{e}^{i \parallel} \ud t
+
\underline{a}_i^\parallel(t),
\end{multline}
where the notation $()^{-1}_\perp$ means a projection to the subspace where the matrix $\widetilde{E}^i$ is invertible followed by an inversion in that subspace.
This equation is our general result for the timestep of the gauge field fluctuation.

In the small $\ud t$ limit 
$\widetilde{\Box}_{0i} \approx \mathbb{1} + i \widetilde{E}^i \ud t$ 
and we see that \eq\nr{eq:finalastep} reduces to $\underline{a}_i(t+\ud t) = \underline{e}^i \ud t + \underline{a}_i(t)$ as desired. It may seem like a disproportionate amount of trouble to formulate the equation in this way, when the result reduces to the naive discretization in the limit $\ud t\to 0$ which one wants to take in the end. However, we have found that in practical computations it is essential for a good precision to conserve Gauss's law also in discrete time and not only in the continuous time limit. At this point it is also straightforward to check that the equation  is time reversal invariant, ensuring second order accuracy in $\ud t$.

Note that the form~\nr{eq:finalastep} of the timestep results from a choice made in writing the timestep for $e^i$ and Gauss's law in the form \eqs\nr{eq:estep}, \nr{eq:gauss}. We could have resolved the ambiguity in linearizing the fluctuations of the timelike plaquette in another way by defining a different electric field fluctuation e.g. by
\begin{equation}
 \underline{e}^{i}_{\textup{mod}} = 
\left(\ud t \widetilde{E}^i\right)^{-1}_\perp \left[
i \left( \widetilde{\Box}^\dag_{0i} - \mathbb{1} \right) \underline{e}^{i \perp}
\right]
+ 
\underline{e}^{i \parallel}.
\end{equation}

This would make the timestep for $a_i$ simpler, but the timestep and Gauss's law for $e^{i}_\textup{mod}$ would have a more complicated form, with the appearence of  terms proportional to $\widetilde{E}^i \ud t$.

The general result~\nr{eq:finalastep} requires the solution of a system of $\nc^2-1$ linear equations. For the special case of SU(2) we can invert the matrix $\widetilde{E}^i$ analytically using the fact that in the absence of symmetric structure constants the Fierz identity for $f^{abc}f^{ade}$ is particularly simple $\epsilon^{ijk}\epsilon^{ilm} = \delta^{jl}\delta^{km}-
\delta^{jm}\delta^{kl}$. Thus if, for the perpendicular part, $E^i_a a^\perp_{i,a}=0$, we have $\widetilde{E}^i\widetilde{E}^i \underline{a}_i^\perp = E^i_a E^i_a \underline{a}_i^\perp$, and we can write
\nr{eq:finalastep} as
\begin{multline}\label{eq:su2adj}
 \underline{a}_i(t+\ud t) = \frac{1}{E^i_a E^i_a} \widetilde{E}^i \Bigg\{ \left[
-i \left( \widetilde{\Box}_{0i} - \mathbb{1} \right) \underline{e}^{i \perp}
+ \widetilde{E}^i  \widetilde{\Box}_{0i}\underline{a}_i^\perp(t)
\right]
\\
+ 
\underline{e}^{i \perp} \ud t
+
\underline{a}_i^\perp(t) \Bigg\}, 
\end{multline}
or in the fundamental representation as
\begin{multline}\label{eq:su2fund}
a_i(t+\ud t) = \frac{i}{2\tr \left[E^i E^i\right]} 
\Bigg[E^i,
\\
-i \left(\Box_{0i}e^{i \perp} \Box^\dag_{0i} - e^{i \perp}\right)
+
\left[E^i,\Box_{0i} a^\perp_i(t)\Box^\dag_{0i}\right]
\Bigg]
\\
+ \ud t e^{i \parallel} + a^\parallel_i(t)
\end{multline}
We stress that these final versions \nr{eq:su2adj} and~\nr{eq:su2fund} are valid for SU(2) only, and e.g. for SU(3) one must use \eq\nr{eq:finalastep}.

\section{Numerical tests}
\label{sec:num}

\begin{figure}[tb!]
\centerline{\includegraphics[width=0.48\textwidth]{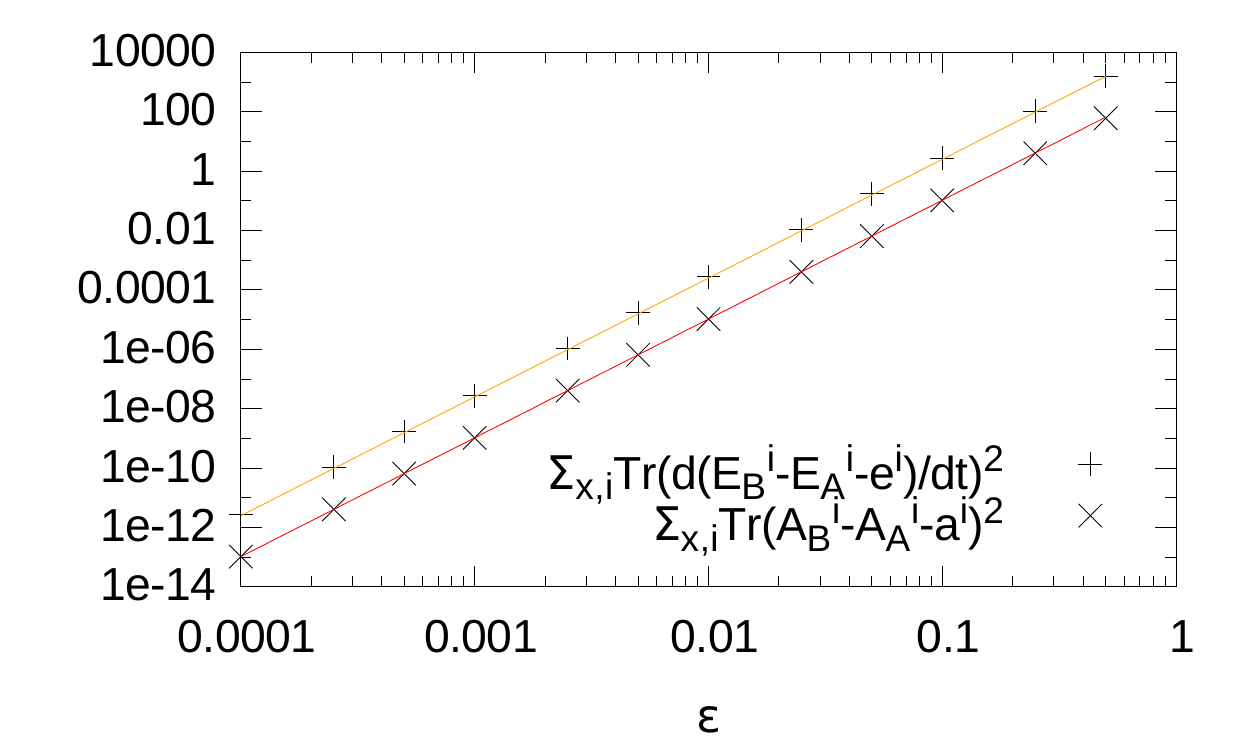}}
\caption{Test of the decomposition of the field in the background field and fluctuation after some finite time. All runs have been evolved to same physical time, which is $\ud t N_t = 2$ here with $\ud t=0.01$ and $N_t = 200.$ The upper points correspond to $\delta_{\dot E}$ and the lower $\delta_A$ (see Eqs~(\ref{eq:de}) and (\ref{eq:da})).
The straight lines are fits of the form $a x^4,$ showing that the observables decrease with the correct power law.
}
\label{fig:a_t_comparison}
\end{figure}

We now have the equations of motion for the linearized fluctuation: the timestep for $e^i$ (\eq\nr{eq:estep}), for $a_i$ (general equation in 
\eq\nr{eq:finalastep} and SU(2)-specific ones in \eqs\nr{eq:su2adj} and~\nr{eq:su2fund}) and Gauss's law~\nr{eq:gauss}.  We present here some simple test results from an implementaion of  these equations for the SU(2) gauge group.

We construct initial conditions for a background field configuration by setting the gauge fields $A_i$ to random values uniformly distributed in the interval $[0,0.9]$. The electric fields are set to zero initially in order to satisfy the Gauss's law $C(\xx,0)=0$. We then  construct the link matrices by exponentiating the gauge fields $U_i = e^{iag A_i}$. We similarly construct initial fluctuation fields. We then choose a small parameter $\varepsilon$ ranging from 0.5 to 0.0001 and multiply the fluctuations by $\varepsilon$,  effectively setting $\varepsilon$ as the scale of these fluctuations, i.e. $e^i, a_i \sim \varepsilon$. 
We can now evolve separately in time: 
\begin{enumerate}
\item The system of the  background field and linearized fluctuations $E^i, A_i, e^i, a_i$ and
\item  A different pure background field configuration initialized as $\hat E^i(t=0) = E^i(t=0) + e^i(t=0)$ and
$\hat A^i(t=0) = A^i(t=0) + a^i(t=0)$. 
\end{enumerate}

\begin{figure}[tb!]
\centerline{\includegraphics[width=0.48\textwidth]{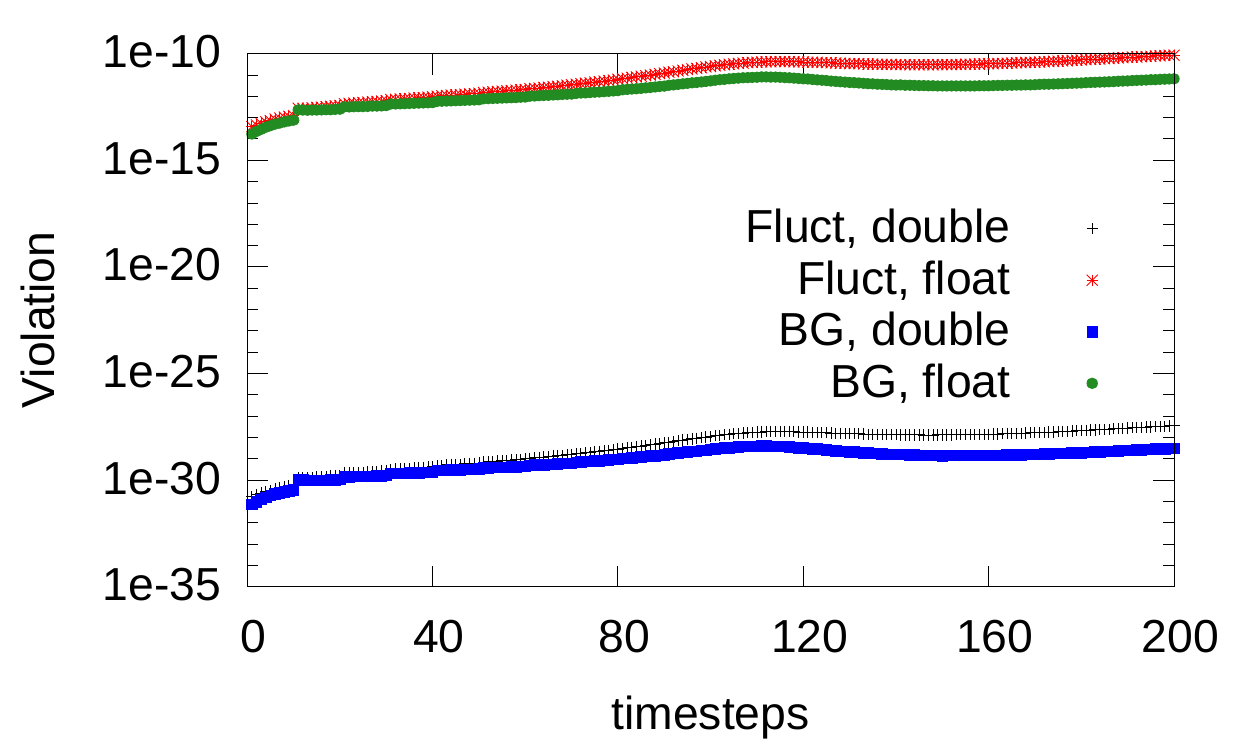}}
\caption{Violation of Gauss's law as a function of time, in single and double precision for the background field only compared to the fluctuations. We have performed a random gauge transformation on every timestep and fixed Coulomb gauge on every tenth timestep to verify the gauge invariance also numerically. Here $\varepsilon=0.1$ and $a_t=0.01.$ The expressions used to measure the violations are given by equations (\ref{eq:bggaussviol}) and (\ref{eq:flucgaussviol}).}
\label{fig:gausslaw}
\end{figure}

If we have now successfully linearized the classical equations of motion, the squared differences 
\begin{equation}
\label{eq:de}
\delta_E = \sum \limits_{x,i}\tr(\hat E^i-E^i-e^i)^2
\end{equation}
 and 
 \begin{align} 
\label{eq:da}
 \delta_A & = \dfrac{1}{2} \sum \limits_{x,i,a}\left(2 {\rm Im}\mathrm{Tr}\left( t^a \hat U_{i}U_{i}^\dagger\right) - a_i^a\right)^2\\
  &\approx\sum \limits_{x,i} \tr(\hat A^i-A^i-a^i)^2
 \end{align} 
 should scale as $\varepsilon^4$ with the magnitude of the fluctuation. For numerical convenience it is easier for us to plot the corresponding differences for the time derivatives $E^i(t+\ud t) - E^i(t)$ etc. as the expression involving time derivatives is easily obtained during the time-evolution.
 In \fig~\ref{fig:a_t_comparison} we show that indeed these differences scale in the correct way for a large range of $\varepsilon$. We can note here that for a naive $a_i$ timestep $a_i(t+\ud t)=  a_i(t)+ e^i$ the correct $\varepsilon$-scaling for small fluctuations is only obtained for prohibitively expensive small values of $\ud t$ because the correct scaling of $\delta_E$ and $\delta_A$ is violated by terms of the order $\varepsilon^2 \ud t^4$.  We have verified this numerically by observing that  for larger $\ud t$ $\delta_E$ and $\delta_A$ begin to scale as  $\varepsilon^2$. This means that the naive timestep for $a_i$ does not correctly capture the difference $\hat A^i-A^i$ even to leading order in $\varepsilon$.

We also show, in \fig\ref{fig:gausslaw}, the violation of Gauss's law constraint as a function of time. To quantify the violation we consider for the background field
\begin{widetext}
\begin{equation} \label{eq:bggaussviol}
\frac{2\sum \limits_x \mathrm{Tr}\left(\sum \limits_{i}\left[E^i(\xx) - U^\dag_i(\xx-\ii)E^i(\xx-\ii)U_i(\xx-\ii)\right]\right)^2}{2 \sum \limits_{x,i} \mathrm{Tr}\left[E^i(\xx) - U^\dag_i(\xx-\ii)E^i(\xx-\ii)U_i(\xx-\ii)\right]^2},
\end{equation}
and similarly for the fluctuations
\begin{equation} \label{eq:flucgaussviol}
 \frac{2 \sum \limits_x \mathrm{Tr}\left(\sum \limits_i \left[
e^i(\xx)- U_i^\dagger(\xx-\ii) e^i(\xx-\ii) U_i(\xx-\ii)
 + i U_i^\dagger(\xx-\ii)[a_i(\xx-\ii),E^i(\xx-\ii)]U_i(\xx-\ii)
\right]\right)^2 
}{
2 \sum \limits_{x,i} \mathrm{Tr} \left[
e^i(\xx)- U_i^\dagger(\xx-\ii) e^i(\xx-\ii) U_i(\xx-\ii)
 + i U_i^\dagger(\xx-\ii)[a_i(\xx-\ii),E^i(\xx-\ii)]U_i(\xx-\ii)
\right]^2}.
\end{equation} 
\end{widetext}
These two quantities measure how well the components in different spatial  directions cancel each other. Due to numerical roundoff error, Gauss's law is never satisfied exactly. However, our algorithm preserves it equally well for the fluctuations as for the background field. Also, the fact that Gauss's law remains satisfied orders of magnitude more precisely in double than single precision shows that that the remaining values are purely due to the limited machine precision.

\section{Conclusions and outlook}
\label{sec:conc}

Unstable fluctuations seeded by quantum effects around a boost invariant classical background field play an important part in the pre-equilibrium evolution of heavy-ion collisions. 
Until now, the Classical Statistical Approximation has been common tool to study these phenomena. However, the very UV dominated spectrum of vacuum fluctuations in field theory makes attaining the continuum limit in CSA calculations very difficult if not impossible.

We have argued in this paper that it would be desirable to address these issues by real time lattice calculations with an explicitly linearized fluctuation around the classical field. We have here explicitly derived and tested equations of motion for these fluctuations, showing that satisfying Gauss's law for the fluctuations requires a careful treatment in the discretization of the timestep.
By giving up the attempt to resum asymptotical long time ``secular divergences,'' which are not a problem with a matching to kinetic theory, one stands to gain better control of the UV dynamics in the classical gauge field calculation. 
We expect this formalism to have several interesting applications, which we plan to return to in future work.

\begin{acknowledgments}
  We thank K. Boguslavski, S. Schlichting and Y. Zhu for discussions.
  T.~L.\ is supported by the Academy of Finland, projects 267321,
  273464  and 303756,  and J.P. by the Jenny and Antti Wihuri Foundation.
\end{acknowledgments}

\bibliography{spires}
\bibliographystyle{JHEP-2modlong}

\end{document}